\documentclass{jpsj-suppl}
\usepackage{txfonts} 
\usepackage{bm}

\title{
A Fast Algorithm for Lattice Hyperonic Potentials 
}

\author{
Hidekatsu \textsc{Nemura},$^{1,2}$ %
Sinya~\textsc{Aoki},$^{1,2,3}$
Takumi~\textsc{Doi},$^2$ 
Shinya~\textsc{Gongyo},$^{2,3}$
Tetsuo~\textsc{Hatsuda},$^{2,4}$
Yoichi~\textsc{Ikeda},$^2$
Takashi~\textsc{Inoue},$^{2,5}$
Takumi~\textsc{Iritani},$^{2,6}$
Noriyoshi~\textsc{Ishii},$^{2,7}$
Takaya~\textsc{Miyamoto},$^{2,3}$
Keiko~\textsc{Murano},$^{2,7}$
and
Kenji~\textsc{Sasaki}$^{1,2}$
}

\inst{
$^{1}$
Center for Computational Sciences, University of Tsukuba, %
Ibaraki, 305-8577, Japan 
\\
$^{2}$
Theoretical Research Division, Nishina Center, RIKEN, %
Saitama, 351-0198, Japan
\\
$^{3}$
Yukawa Institute for Theoretical Physics, Kyoto University, %
Kyoto, 606-8502, Japan
\\
$^{4}$
Kavli IPMU (WPI), The University of Tokyo, %
Chiba 277-8583, Japan
\\
$^{5}$
Nihon University, College of Bioresource Sciences, 
Kanagawa 252-0880, Japan
\\
$^{6}$
Department of Physics and Astronomy, Stony Brook University, Stony Brook, 
New York 11794-3800, USA
\\
$^{7}$
Research Center for Nuclear Physics (RCNP), Osaka University, 
Osaka 567-0047, Japan
}

\email{nemura.hidekatsu.gb@u.tsukuba.ac.jp}

\abst{
We describe an 
efficient 
algorithm to compute a large number of 
baryon-baryon interactions from $NN$ to $\Xi\Xi$ 
by means of HAL QCD method, which lays the groundwork for the 
nearly physical point lattice QCD calculation with 
volume 
$(96a)^4\approx$($8.2$fm)$^4$. 
Preliminary results of $\Lambda N$ potential calculated with 
quark masses corresponding to ($m_{\pi}$,$m_{K}$)$\approx$(146,525)MeV 
are presented.

}

\kword{
nuclear force, hyperon-nucleon interaction, lattice QCD, high performance computing
}

\begin{document}
\hfill~UTCCS-P-89, RIKEN-QHP-230, YITP-16-67
\vspace*{-1.2em}
\maketitle

\section{Introduction}

Nuclear force and hyperonic nuclear forces provide 
a useful starting point 
to figure out how the hypernuclear systems are bound, 
in which hyperons (or strange quarks) are embedded 
in normal nuclear systems as ``impurities.''\cite{Hashimoto:2006aw} 
From studies of few-body systems for $s$-shell $\Lambda$ hypernuclei, 
for example, 
it has been pointed out that the coupled-channel $\Lambda N-\Sigma N$ 
interaction plays a vital role to make a hypernucleus being 
bounded\cite{Nemura:2002fu} 
although phenomenological hyperon-nucleon potentials are not well 
constrained 
from experimental data. 
In the past several years, 
a new approach to study the hadronic forces from the lattice QCD 
has been proposed\cite{Ishii:2006ec,Aoki:2009ji}. 
In this approach, the interhadron potential is obtained 
by means of the lattice QCD measurement of the Nambu-Bethe-Salpeter (NBS) 
wave function. 
The observables such as the phase shifts and the binding energies are 
calculated through the resultant potential\cite{Aoki:2012tk}. 
This approach has been further extended and applied to various problems. 
See Ref.\cite{Sasaki:2015ifa} and references therein for the 
state-of-the-art outcomes. 
In addition, a large scale lattice QCD calculation is 
now in progress\cite{DoiIshiiSasaki2015HYP} to study the baryon 
interactions from $NN$ to $\Xi\Xi$ 
by measuring 
the NBS wave functions for 
52 channels from the $2+1$ flavor lattice QCD, 
which is 
founded on inconspicuous but vital work\cite{Nemura:2015yha} that 
reveals 
a beneficial algorithm for computing a large number of 
NBS wave functions simultaneously; 
see Eqs.~(34)-(38) in Ref.~\cite{Nemura:2015yha} for the specific 
channels of the above 52 NBS wave functions. 
The purpose of this report is to present 
a key aspect of 
the algorithm that performs efficiently a concurrent computation of 
such a lot of NBS wave functions for various baryon channels. 
As a preliminary snapshot of the ongoing work, 
effective $\Lambda N$ potential 
at almost physical quark masses 
corresponding to ($m_{\pi}$,$m_{K}$)$\approx$(146,525)MeV is 
also presented, which is obtained from the single correlator 
$\langle p\Lambda\overline{p\Lambda}\rangle$ 
by adopting a recipe 
in which the effects from the $\Sigma N$ channel are 
effectively included. 

\section{Effective Baryon Block Algorithm}

In the HAL QCD method, the 
interaction is obtained 
through the four-point 
correlator 
defined by
\begin{equation}
\begin{array}{c}
 { F}_{\alpha_{1}\alpha_{2},\alpha_{3}\alpha_{4}}
     ^{\langle B_1B_2\overline{B_3B_4}\rangle}(\vec{r},t-t_0) = 
 \sum_{\vec{X}}
 \left\langle 0
  \left|
   B_{1,{\alpha_1}}(\vec{X}+\vec{r},t)
   B_{2,{\alpha_2}}(\vec{X},t)
   \overline{{\cal J}_{B_{3,\alpha_{3}} B_{4,\alpha_{4}}}
                     (t_0)}
  \right| 0 
 \right\rangle,
\end{array}
\end{equation}
where the summation over $\vec{X}$ selects  states with 
zero total momentum. 
The $B_{1,\alpha_1}(x)$ and $B_{2,\alpha_2}(y)$ denote the 
interpolating fields of the baryons such as 
%
\begin{equation}
 \!\!\!
 \begin{array}{llll}
  p \! = \! \varepsilon_{abc} \left(
			 u_a C\gamma_5 d_b
			\right) u_c,\!
  &
  n \! = \! - \varepsilon_{abc} \left(
			   u_a C\gamma_5 d_b
			  \right) d_c,\!
  &
  \Sigma^{+} \! = \! - \varepsilon_{abc} \left(
				    u_a C\gamma_5 s_b
				   \right) u_c,\!
  &
  \Sigma^{-} \! = \! - \varepsilon_{abc} \left(
				    d_a C\gamma_5 s_b
				   \right) d_c,\!
  \\
  \Sigma^{0} \! = \! {1\over\sqrt{2}} \left( X_u \! - \! X_d \right),\!
  &
  \Lambda \! = \! {1\over \sqrt{6}} \left( X_u \! + \! X_d \! - \! 2 X_s \right),\!
  &
  \Xi^{0} \! = \! \varepsilon_{abc} \left(
                               u_a C\gamma_5 s_b
                              \right) s_{c},\!

  &
  \Xi^{-} \! = \! - \varepsilon_{abc} \left(
                                 d_a C\gamma_5 s_b
                                \right) s_{c},\!
  \\
  \mbox{where}
  &
  X_u = \varepsilon_{abc} \left( d_a C\gamma_5 s_b \right) u_c, 
  &
  X_d = \varepsilon_{abc} \left( s_a C\gamma_5 u_b \right) d_c,
  &
  X_s = \varepsilon_{abc} \left( u_a C\gamma_5 d_b \right) s_c.
 \end{array}
 \label{BaryonOperatorsOctet}
\end{equation}
%
For simplicity, we have suppressed the explicit spinor indices and 
spatial coordinates in Eq.~(\ref{BaryonOperatorsOctet}). 
%
$\overline{{\cal J}_{B_{3,\alpha_{3}}B_{4,\alpha_{4}}}
                   (t_0)}$ 
is the source operator which creates 
$B_{3}B_{4}$ states. 
Hereafter, the explicit time dependence is suppressed for simplicity.
The 
4pt correlator 
is evaluated through considering the Wick's contraction 
together with defining the baryon blocks 
$[B_{1,\alpha_{1}}^{(0)}](\vec{x};~\bm{\xi}_{P_{1}P_{2}P_{3}}^\prime)$ and 
$[B_{2,\alpha_{2}}^{(0)}](\vec{y};~\bm{\xi}_{P_{4}P_{5}P_{6}}^\prime)$, 
%
\begin{eqnarray}
 { F}
     ^{\langle B_1B_2\overline{B_3B_4}\rangle}
     _{{\alpha_1}{\alpha_2},{\alpha_3}{\alpha_4}}(\vec{r}%
                                                          )
 &=& 
  \sum_{\vec{X}} 
   \sum_{P} \sigma_{P}
   ~
   [B_{1,\alpha_1}^{(0)}](\vec{X}+\vec{r};
   ~
   \xi^{\prime}_{P_{1}}, \xi^{\prime}_{P_{2}}, \xi^{\prime}_{P_{3}}
                         ) 
   ~
   [B_{2,\alpha_2}^{(0)}](\vec{X};
   ~
   \xi^{\prime}_{P_{4}}, \xi^{\prime}_{P_{5}}, \xi^{\prime}_{P_{6}}
                         )
  \nonumber 
 \\
 &&
 \qquad 
 \times 
 \varepsilon_{c_{{1}}^{\prime} c_{{2}}^{\prime} c_{{3}}^{\prime}}
 \varepsilon_{c_{{4}}^{\prime} c_{{5}}^{\prime} c_{{6}}^{\prime}}
 (C\gamma_5)_{\alpha_{1}^{\prime}\alpha_{2}^{\prime}}
 (C\gamma_5)_{\alpha_{4}^{\prime}\alpha_{5}^{\prime}}
 \delta_{\alpha_{3}^{\prime}\alpha_{3}}
 \delta_{\alpha_{6}^{\prime}\alpha_{4}},
 \label{NaiveBaryonBlocks}
\end{eqnarray}
\begin{equation}
 \begin{array}{ll}
  \!\!\!\mbox{with} &
  ~[B_{1,\alpha_{1}}^{(0)}](\vec{x};~
        \bm{\xi}_{P_{1}P_{2}P_{3}}^\prime) 
  =
  [B_{1,\alpha_{1}}^{(0)}](\vec{x};~
   \xi_{P_{1}}^{\prime}, \xi_{P_{2}}^{\prime}, \xi_{P_{3}}^{\prime}
                        ) 
  =
  \left\langle 
   B_{1,\alpha_{1}}(\vec{x})
   ~
   \bar{q}_{B_{1},3}^{\prime}(\xi_{P_{3}}^{\prime})
   \bar{q}_{B_{1},2}^{\prime}(\xi_{P_{2}}^{\prime})
   \bar{q}_{B_{1},1}^{\prime}(\xi_{P_{1}}^{\prime})
 \right\rangle,
 \quad \mbox{and} 
 \\
 &
 ~
 [B_{2,\alpha_{2}}^{(0)}](\vec{y};~
  \bm{\xi}_{P_{4}P_{5}P_{6}}^\prime) 
  =
  [B_{2,\alpha_{2}}^{(0)}](\vec{y};~
   \xi_{P_{4}}^{\prime}, \xi_{P_{5}}^{\prime}, \xi_{P_{6}}^{\prime}
                        ) 
  =
  \left\langle 
   B_{2,\alpha_{2}}(\vec{y})
   ~
   \bar{q}_{B_{2},6}^{\prime}(\xi_{P_{6}}^{\prime})
   \bar{q}_{B_{2},5}^{\prime}(\xi_{P_{5}}^{\prime})
   \bar{q}_{B_{2},4}^{\prime}(\xi_{P_{4}}^{\prime})
   \right\rangle,
 \end{array}
\end{equation}
%
where $\sigma_{P}$ and 
$\{\xi_{P_{1}}^{\prime},\cdots,\xi_{P_{6}}^{\prime}\}$ 
are the sign factor and the set of permutated 
spin-color-space-time coordinates 
for each permutation $P$ due to the Wick's contraction, respectively. 
Both 3-tuple sets of the quark fields 
$\{\bar{q}_{B_{1},1}^{\prime},\bar{q}_{B_{1},2}^{\prime},
   \bar{q}_{B_{1},3}^{\prime}\}$ and 
$\{\bar{q}_{B_{2},4}^{\prime},\bar{q}_{B_{2},5}^{\prime},
   \bar{q}_{B_{2},6}^{\prime}\}$ 
are ordered 
so as to create 
the $B_{1}$ and $B_{2}$ states properly. 
Taking the expression in Eq.~(\ref{NaiveBaryonBlocks}), 
the number of the iterations to obtain a $
 { F}
     ^{\langle B_1B_2\overline{B_3B_4}\rangle}
     _{{\alpha_1}{\alpha_2},{\alpha_3}{\alpha_4}}(\vec{r}%
                                                          )$ 
except the spatial degrees of freedom 
reduces to 
$(N_{c}!N_{\alpha})^{B}\times 
N_{u}!N_{d}!N_{s}!\times 
2^{N_{\Lambda}+N_{\Sigma^{0}}-B}$ 
from the number of iterations in naive counting that is 
$(N_{c}!N_{\alpha})^{2B}\times N_{u}!N_{d}!N_{s}!$, 
where $N_{c}=3, N_{\alpha}=4$ and 
$N_{\Lambda}$, $N_{\Sigma^{0}}$, 
$N_{u},N_{d},N_{s}$ and $B$ are the numbers of 
$\Lambda,\Sigma^{0}$, up-quark, down-quark, strange-quark and the baryons 
(i.e., always $B=2$ in the present study), respectively. 
Through the employment of Fast-Fourier-Transform 
(FFT)\cite{Ishii:2009zr,Nemura:2009kc} 
we attain the expression in terms of the 
effective baryon blocks\cite{Nemura:2015yha} 
%
\begin{equation}
 { F}
     ^{\langle B_1B_2\overline{B_3B_4}\rangle}
     _{{\alpha_1}{\alpha_2},{\alpha_3}{\alpha_4}}\!(\vec{r}%
                                                          )
\! =\!\!\!
  \sum_{P} \! \sigma_{P} \!\!
   \sum_{\vec{X}} 
   \! \left( 
         [\!B_{1,\alpha_1}^{(P)}\!](\vec{X}\!\!+\!\!\vec{r}) \! \times \!
         [B_{2,\alpha_2}^{(P)}](\vec{X})
         \right)_{\alpha_3\alpha_4%
                                   }
         \nonumber
\!\! =\!\!
  {1\over L^3}
   \!\! \sum_{\vec{q}}
    \left(
     \! \sum_{P} \! \sigma_{P} \!
      \left(
           [\widetilde{B_{1,\alpha_1}^{(P)}}]( \vec{q}) \! \times \!
           [\widetilde{B_{2,\alpha_2}^{(P)}}](-\vec{q})
      \! \right)_{\alpha_3\alpha_4}
     \! \right)
     \! {\rm e}^{i\vec{q}\cdot\vec{r}}.
  \label{B1B2.B1B2B4B3.FFT}
\end{equation}
%
The effective baryon blocks, $
      \left(
           [\widetilde{B_{1,\alpha_1}^{(P)}}]( \vec{q}) \times 
           [\widetilde{B_{2,\alpha_2}^{(P)}}](-\vec{q})
      \right)_{\alpha_{3}\alpha_{4}},
$
are obtained by multiplying tensorial factors with the 
normal baryon blocks; 
for example, 
the specific form of the 
4pt correlator 
${F}^{\langle p \Lambda\overline{pX_{u}}\rangle}
    _{{\alpha_1}{\alpha_2},{\alpha_{3}}{\alpha_{4}}}(\vec{r})$ 
of the $\langle p \Lambda\overline{pX_{u}}\rangle$ channel 
is given by~\cite{Nemura:2015yha},
\begin{eqnarray}
 {F}^{\langle p \Lambda\overline{pX_{u}}\rangle}
    _{{\alpha_1}{\alpha_2},{\alpha_3}{\alpha_4}}(\vec{r})
 &=&
  {1\over L^3}
   \sum_{\vec{q}}
    \left(
%
    [\widetilde{      p}_{\alpha_1\alpha_3}^{(1)}]( \vec{q}) 
    [\widetilde{\Lambda}_{\alpha_2\alpha_4}^{(1)}](-\vec{q})
    -
    [\widetilde{      p}_{\alpha_1\alpha_4}^{(2)}]_{c_{3}^{\prime},c_{6}^{\prime}}( \vec{q}) 
    [\widetilde{\Lambda}_{\alpha_2\alpha_3}^{(2)}]_{c_{3}^{\prime},c_{6}^{\prime}}(-\vec{q})
    \right.
    \nonumber
    \\
    &&
    \left.
    -
    [\widetilde{      p}_{\alpha_1\alpha_3}^{(3)}]_{c_{2}^{\prime},\alpha_{2}^{\prime},c_{4}^{\prime},\alpha_{4}^{\prime}}( \vec{q}) 
    [\widetilde{\Lambda}_{\alpha_2\alpha_4}^{(3)}]_{c_{2}^{\prime},\alpha_{2}^{\prime},c_{4}^{\prime},\alpha_{4}^{\prime}}(-\vec{q})
    \right.
    \left.
    +
    [\widetilde{      p}_{\alpha_1\alpha_4}^{(4)}]_{c_{1}^{\prime},\alpha_{1}^{\prime},c_{5}^{\prime},\alpha_{5}^{\prime}}( \vec{q}) 
    [\widetilde{\Lambda}_{\alpha_2\alpha_3}^{(4)}]_{c_{1}^{\prime},\alpha_{1}^{\prime},c_{5}^{\prime},\alpha_{5}^{\prime}}(-\vec{q})
    \right.
    \nonumber
    \\
    &&
    \left.
    +
    [\widetilde{      p}_{\alpha_1\alpha_3\alpha_4}^{(5)}]_{c_{1}^{\prime},\alpha_{1}^{\prime},c_{6}^{\prime}}( \vec{q}) 
    [\widetilde{\Lambda}_{\alpha_2                }^{(5)}]_{c_{1}^{\prime},\alpha_{1}^{\prime},c_{6}^{\prime}}(-\vec{q})
    \right.
    \left.
    -
    [\widetilde{      p}_{\alpha_1\alpha_3\alpha_4}^{(6)}]_{c_{3}^{\prime},c_{5}^{\prime},\alpha_{5}^{\prime}}( \vec{q}) 
    [\widetilde{\Lambda}_{\alpha_2                }^{(6)}]_{c_{3}^{\prime},c_{5}^{\prime},\alpha_{5}^{\prime}}(-\vec{q})
    \right)
    {\rm e}^{i\vec{q}\cdot\vec{r}},
  \label{pL.pLXup.FFT}
\end{eqnarray}
where
\begin{eqnarray}
 &&
  \begin{array}{l}
   ~[\widetilde{      p}_{\alpha_1\alpha_3}^{(1)}]( \vec{q}) 
    =
    [\widetilde{      p}_{\alpha_{1}}^{(0)}](\vec{q};~\bm{\xi}_{{1}{2}{3}}^\prime) 
    \varepsilon_{c_{{1}}^{\prime} c_{{2}}^{\prime} c_{{3}}^{\prime}} (C\gamma_5)_{\alpha_{1}^{\prime}\alpha_{2}^{\prime}} \delta_{\alpha_{3}^{\prime}\alpha_{3}},
   \\
   ~[\widetilde{\Lambda}_{\alpha_2\alpha_4}^{(1)}](-\vec{q})
   =
   [\widetilde{\Lambda}_{\alpha_{2}}^{(0)}](-\vec{q};\bm{\xi}_{{4}{5}{6}}^\prime) 
   \varepsilon_{c_{{4}}^{\prime} c_{{5}}^{\prime} c_{{6}}^{\prime}} (C\gamma_5)_{\alpha_{4}^{\prime}\alpha_{5}^{\prime}} \delta_{\alpha_{6}^{\prime}\alpha_{4}},
  \end{array}
  \\
 &&
  \begin{array}{l}
   ~[\widetilde{      p}_{\alpha_1\alpha_4}^{(2)}]_{c_{3}^{\prime}c_{6}^{\prime}}( \vec{q}) 
    =
    [\widetilde{      p}_{\alpha_{1}}^{(0)}](\vec{q};~\bm{\xi}_{{1}{2}{6}}^\prime) 
    \varepsilon_{c_{{1}}^{\prime} c_{{2}}^{\prime} c_{{3}}^{\prime}} (C\gamma_5)_{\alpha_{1}^{\prime}\alpha_{2}^{\prime}} \delta_{\alpha_{6}^{\prime}\alpha_{4}},
   \\
   ~[\widetilde{\Lambda}_{\alpha_2\alpha_3}^{(2)}]_{c_{3}^{\prime}c_{6}^{\prime}}(-\vec{q})
   =
   [\widetilde{\Lambda}_{\alpha_{2}}^{(0)}](-\vec{q};\bm{\xi}_{{4}{5}{3}}^\prime) 
   \varepsilon_{c_{{4}}^{\prime} c_{{5}}^{\prime} c_{{6}}^{\prime}} (C\gamma_5)_{\alpha_{4}^{\prime}\alpha_{5}^{\prime}} \delta_{\alpha_{3}^{\prime}\alpha_{3}},
  \end{array}
  \\
 &&
  \begin{array}{l}
   ~[\widetilde{      p}_{\alpha_1\alpha_3}^{(3)}]_{c_{2}^{\prime}\alpha_{2}^{\prime}c_{4}^{\prime}\alpha_{4}^{\prime}}( \vec{q}) 
    =
    [\widetilde{      p}_{\alpha_{1}}^{(0)}](\vec{q};~\bm{\xi}_{{1}{4}{3}}^\prime) 
    \varepsilon_{c_{{1}}^{\prime} c_{{2}}^{\prime} c_{{3}}^{\prime}} (C\gamma_5)_{\alpha_{1}^{\prime}\alpha_{2}^{\prime}} \delta_{\alpha_{3}^{\prime}\alpha_{3}},
   \\
   ~[\widetilde{\Lambda}_{\alpha_2\alpha_4}^{(3)}]_{c_{2}^{\prime}\alpha_{2}^{\prime}c_{4}^{\prime}\alpha_{4}^{\prime}}(-\vec{q})
   =
   [\widetilde{\Lambda}_{\alpha_{2}}^{(0)}](-\vec{q};\bm{\xi}_{{2}{5}{6}}^\prime) 
   \varepsilon_{c_{{4}}^{\prime} c_{{5}}^{\prime} c_{{6}}^{\prime}} (C\gamma_5)_{\alpha_{4}^{\prime}\alpha_{5}^{\prime}} \delta_{\alpha_{6}^{\prime}\alpha_{4}},
  \end{array}
  \\
 &&
  \begin{array}{l}
   ~[\widetilde{      p}_{\alpha_1\alpha_4}^{(4)}]_{c_{1}^{\prime}\alpha_{1}^{\prime}c_{5}^{\prime}\alpha_{5}^{\prime}}( \vec{q}) 
    =
    [\widetilde{      p}_{\alpha_{1}}^{(0)}](\vec{q};~\bm{\xi}_{{1}{4}{6}}^\prime) 
    \varepsilon_{c_{{4}}^{\prime} c_{{5}}^{\prime} c_{{6}}^{\prime}} (C\gamma_5)_{\alpha_{4}^{\prime}\alpha_{5}^{\prime}} \delta_{\alpha_{6}^{\prime}\alpha_{4}},
   \\
   ~[\widetilde{\Lambda}_{\alpha_2\alpha_3}^{(4)}]_{c_{1}^{\prime}\alpha_{1}^{\prime}c_{5}^{\prime}\alpha_{5}^{\prime}}(-\vec{q})
   =
   [\widetilde{\Lambda}_{\alpha_{2}}^{(0)}](-\vec{q};\bm{\xi}_{{2}{5}{3}}^\prime) 
   \varepsilon_{c_{{1}}^{\prime} c_{{2}}^{\prime} c_{{3}}^{\prime}} (C\gamma_5)_{\alpha_{1}^{\prime}\alpha_{2}^{\prime}} \delta_{\alpha_{3}^{\prime}\alpha_{3}},
  \end{array}
  \\
 &&
  \begin{array}{l}
   ~[\widetilde{      p}_{\alpha_1\alpha_3\alpha_4}^{(5)}]_{c_{1}^{\prime}\alpha_{1}^{\prime}c_{6}^{\prime}}( \vec{q}) 
    =
    [\widetilde{      p}_{\alpha_{1}}^{(0)}](\vec{q};~\bm{\xi}_{{3}{2}{6}}^\prime) 
    \varepsilon_{c_{{1}}^{\prime} c_{{2}}^{\prime} c_{{3}}^{\prime}} (C\gamma_5)_{\alpha_{1}^{\prime}\alpha_{2}^{\prime}} \delta_{\alpha_{3}^{\prime}\alpha_{3}}
                                                                                                                          \delta_{\alpha_{6}^{\prime}\alpha_{4}},
   \\
   ~[\widetilde{\Lambda}_{\alpha_2                }^{(5)}]_{c_{1}^{\prime}\alpha_{1}^{\prime}c_{6}^{\prime}}(-\vec{q})
    =
    [\widetilde{\Lambda}_{\alpha_{2}}^{(0)}](-\vec{q};\bm{\xi}_{{4}{5}{1}}^\prime) 
    \varepsilon_{c_{{4}}^{\prime} c_{{5}}^{\prime} c_{{6}}^{\prime}} (C\gamma_5)_{\alpha_{4}^{\prime}\alpha_{5}^{\prime}},
  \end{array}
  \\
 &&
  \begin{array}{l}
   ~[\widetilde{      p}_{\alpha_1\alpha_3\alpha_4}^{(6)}]_{c_{3}^{\prime}c_{5}^{\prime}\alpha_{5}^{\prime}}( \vec{q}) 
    =
    [\widetilde{      p}_{\alpha_{1}}^{(0)}](\vec{q};~\bm{\xi}_{{3}{4}{6}}^\prime) 
    \varepsilon_{c_{{4}}^{\prime} c_{{5}}^{\prime} c_{{6}}^{\prime}} (C\gamma_5)_{\alpha_{4}^{\prime}\alpha_{5}^{\prime}}
                                                                                                                         \delta_{\alpha_{3}^{\prime}\alpha_{3}}
                                                                                                                         \delta_{\alpha_{6}^{\prime}\alpha_{4}},
   \\
   ~[\widetilde{\Lambda}_{\alpha_2                }^{(6)}]_{c_{3}^{\prime}c_{5}^{\prime}\alpha_{5}^{\prime}}(-\vec{q})
    =
    [\widetilde{\Lambda}_{\alpha_{2}}^{(0)}](-\vec{q};\bm{\xi}_{{2}{5}{1}}^\prime) 
    \varepsilon_{c_{{1}}^{\prime} c_{{2}}^{\prime} c_{{3}}^{\prime}} (C\gamma_5)_{\alpha_{1}^{\prime}\alpha_{2}^{\prime}},
  \end{array}
  \label{EffectiveBaryonBlocksProtonLambdaStructuredForm}
\end{eqnarray}
%
\begin{eqnarray}
  ~[p_{\alpha_{1}}^{(0)}](\vec{x};~\bm{\xi}_{{1}{2}{3}}^\prime) 
 &=&
 \varepsilon_{b_{{1}} b_{{2}} b_{{3}}} (C\gamma_5)_{\beta_{1}\beta_{2}} \delta_{\beta_{3}\alpha_{1}}
 \det\left|
  \begin{array}{cc}
    \langle u(\zeta_{1}) \bar{u}(\xi_{1}^{\prime}) \rangle &
    \langle u(\zeta_{1}) \bar{u}(\xi_{3}^{\prime}) \rangle
    \\
    \langle u(\zeta_{3}) \bar{u}(\xi_{1}^{\prime}) \rangle &
    \langle u(\zeta_{3}) \bar{u}(\xi_{3}^{\prime}) \rangle
  \end{array}
 \right|
 \langle d(\zeta_{2}) \bar{d}(\xi_{2}^{\prime}) \rangle,
 \\
 ~
 [\Lambda_{\alpha_{2}}^{(0)}](\vec{y};~\bm{\xi}_{{4}{5}{6}}^\prime) 
 &=&
 {1\over \sqrt{6}}
 \varepsilon_{b_{4} b_{5} b_{6}}
  \left\{
   (C\gamma_5)_{\beta_{4}\beta_{5}} \delta_{\beta_{6}\alpha_{2}}
   + (C\gamma_5)_{\beta_{5}\beta_{6}} \delta_{\beta_{4}\alpha_{2}}
   -2 (C\gamma_5)_{\beta_{6}\beta_{4}} \delta_{\beta_{5}\alpha_{2}}
  \right\}
  \nonumber
  \\
  &&
  \qquad
  \times
     {\langle u(\zeta_{6}) \bar{u}(\xi_{6}^{\prime}) \rangle}
      \langle d(\zeta_{4}) \bar{d}(\xi_{4}^{\prime}) \rangle
      \langle s(\zeta_{5}) \bar{s}(\xi_{5}^{\prime}) \rangle.
\end{eqnarray}
%
NB A roll of coordinates, 
$\{\bm{\xi}_{{1}{2}{3}}^\prime, \bm{\xi}_{{4}{5}{6}}^\prime \}$, 
is now called while Ref.~\cite{Nemura:2015yha} has used 
$\{\bm{\xi}_{{1}{4}{2}}^\prime, \bm{\xi}_{{5}{6}{3}}^\prime \}$; 
this is advantageous for generalizing the algorithm toward various 
(e.g., 52) baryon channels. 
By employing the effective block algorithm, 
the number of iterations to evaluate the r.h.s. of Eq.~(\ref{pL.pLXup.FFT})
except the momentum space degrees of freedom becomes 
$1+N_{c}^{2}+N_{c}^{2}N_{\alpha}^{2}+N_{c}^{2}N_{\alpha}^{2}+
N_{c}^{2}N_{\alpha}+N_{c}^{2}N_{\alpha}=370$, 
which is 
remarkably 
smaller than the number 
$(N_{c}!N_{\alpha})^{B}\times 
N_{u}!N_{d}!N_{s}!\times 
2^{N_{\Lambda}+N_{\Sigma^{0}}-B} = 3456$ 
that is 
seen in 
Eq.~(\ref{NaiveBaryonBlocks}). 
The manipulation for the expression of Eq.~(\ref{B1B2.B1B2B4B3.FFT}) 
can be automatically done once the set of the interpolating fields 
(i.e., the quantum numbers) 
of both sink and source parts is given. 

\section{Results}

A large number of baryon-baryon 
potentials 
including the 
channels 
from $NN$ to $\Xi\Xi$ 
are studied through the 
nearly physical point lattice QCD calculation 
by means of HAL QCD method 
with 
$N_{f}=2+1$ dynamical clover fermion gauge configurations 
generated on a $L^4=96^4$ lattice using K computer, 
where the actual computing jobs are launched with the 
unified contraction algorithm (UCA)\cite{Doi:2012xd}; 
see also Ref.\cite{Nemura:2015yha} for the thoroughgoing 
consistency check in the numerical outputs and 
comparison at various occasions 
between the UCA and the present algorithm.
Preliminary studies 
indicate 
that the physical volume is $(aL)^4\approx$(8.2fm)$^4$ with the lattice 
spacing $a\approx 0.085$fm and 
$(m_{\pi},m_{K})\approx(146,525)$MeV. 
See Ref.\cite{Ishikawa:2015rho} for details on the generation of 
the gauge configuration.

\begin{figure}[t]
 \begin{minipage}[t]{0.33\textwidth}
  \centering \leavevmode
  \includegraphics[width=0.99\textwidth]{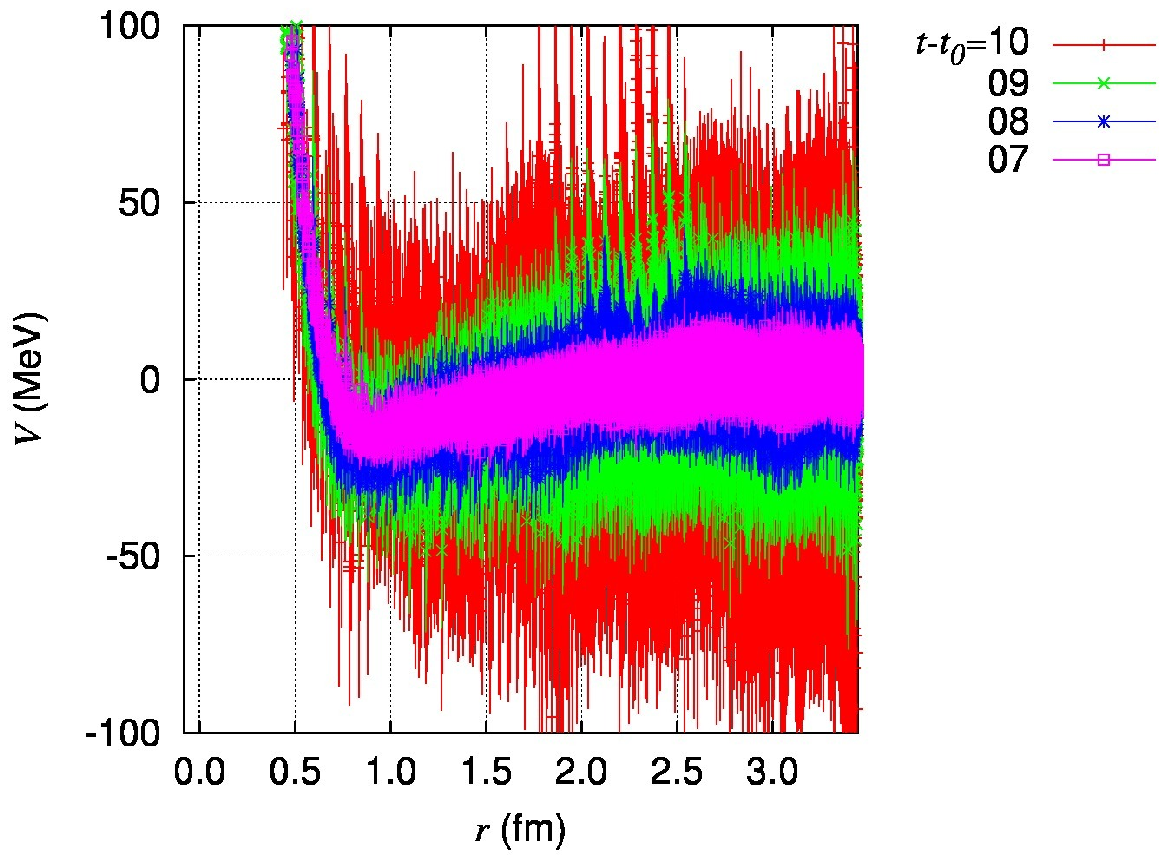}
 \end{minipage}~
 \hfill
 \begin{minipage}[t]{0.33\textwidth}
  \centering \leavevmode
 \includegraphics[width=0.99\textwidth]{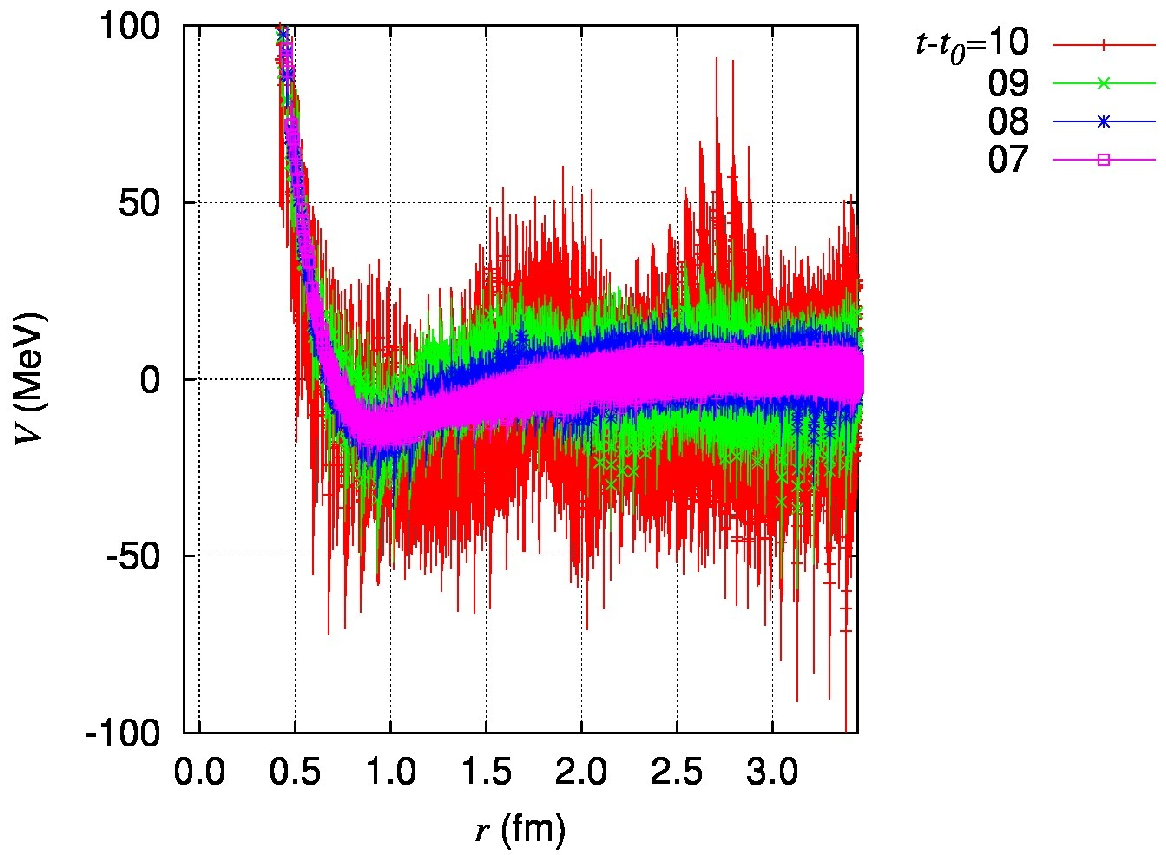}
 \end{minipage}~
 \hfill
 \begin{minipage}[t]{0.33\textwidth}
  \centering \leavevmode
 \includegraphics[width=0.99\textwidth]{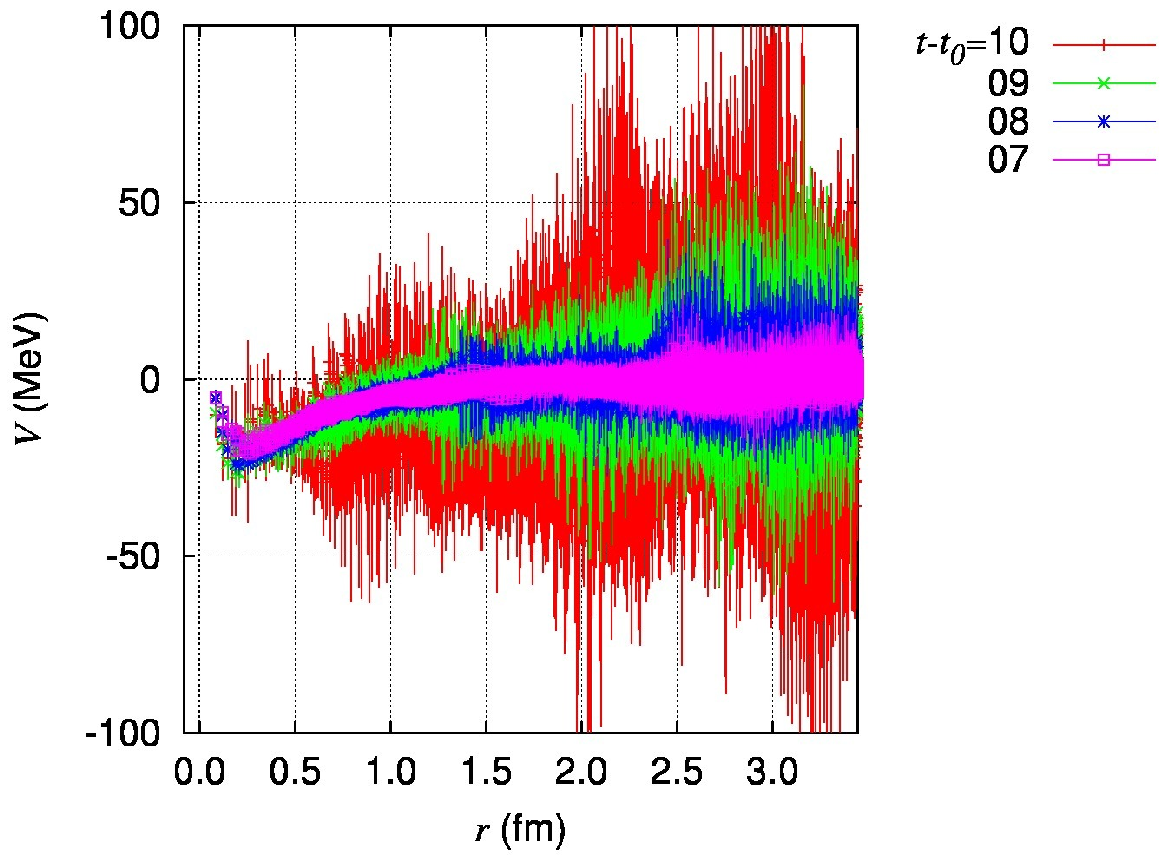}
 \end{minipage}
  \footnotesize
 \caption{Left: $\Lambda N$ central potential in the $^1S_0$ channel 
calculated with nearly physical point 
lattice QCD calculation on a volume $(96a)^4\approx$(8.2fm)$^4$ with the 
lattice spacing $a\approx 0.085$fm and 
$(m_{\pi},m_{K})\approx(146,525)$MeV. 
Center: $\Lambda N$ central potential in the $^3S_1$-$^3D_1$ channel. 
Right: $\Lambda N$ tensor potential in the $^3S_1$-$^3D_1$ channel. 
 \label{Fig_LNpots}}
\end{figure}

Figure~\ref{Fig_LNpots} shows the preliminary snapshots for the 
$\Lambda N$ potential which implicitly includes effects 
from the $\Sigma N$. 
The statistics in the figure is only a fraction about $1/10$ of 
the entire schedule in FY2015. 
The left (center) panel shows the central potential in 
$^1S_0$ ($^3S_1$-$^3D_1$) channel. 
Short-range repulsion and medium-to-long-range 
attraction are found in the range $t-t_{0}=7-9$ for both central potentials. 
The right panel shows the tensor potential in the $^3S_1$-$^3D_1$
channel. 
The results in the range $t-t_{0}=7-9$ show weak tensor force 
in the $\Lambda N-\Lambda N$ diagonal channel.
More comprehensive study which explicitly considers the effects 
from the $\Sigma N$ 
with 
increasing the 
statistics 
will be presented near future. 

%

\section{Summary}

In this paper, we present an efficient algorithm to calculate 
a large number of NBS wave functions simultaneously. 
The effective block algorithm significantly reduces the numerical 
costs for the calculation of 52 two-baryon channels and the 
computer program 
implemented 
lays the foundation for the 
nearly physical point lattice QCD calculation with volume size 
$(96a)^4\approx$($8.2$fm)$^4$. 

The preliminary snapshots of the $\Lambda N$ potential are presented. 
Central potentials in both $^1S_0$ and $^3S_1$-$^3D_1$ channels show the
short-range repulsion and medium-to-long-range attraction. 
The tensor potential of the 
$\Lambda N-\Lambda N$ diagonal channel seems to be weaker than 
the tensor potential of $NN$ interaction. 
Further studies explicitly considering $\Sigma N$ channels 
together with employing larger statistics 
are under progress and will be reported in the near future.

\section*{Acknowledgments}

The lattice QCD calculations have been performed 
on the K computer at RIKEN, AICS 
(Nos. hp120281, hp130023, hp140209, hp150223),
HOKUSAI FX100 computer at RIKEN, Wako (No. G15023)
and HA-PACS at University of Tsukuba (Nos. 
12b-13, 13a-25, 14a-25, 15a-33, 14a-20, 15a-30).
We thank ILDG/JLDG~\cite{ILDGJLDG}
which serves as an essential infrastructure in this study.
This work is supported in part by 
MEXT Grant-in-Aid for Scientific Research 
(25105505, 15K17667, 25287046, 26400281),
and SPIRE (Strategic Program for Innovative REsearch) Field 5 project.
We thank all collaborators in this project.

\end{document}